\documentclass[epj]{webofc}
\usepackage[sort&compress]{natbib} %to cite like [1-4] instead of [1,2,3,4]
\usepackage[varg]{txfonts}   % Web of Conferences font
\usepackage{graphicx}
\usepackage{epstopdf} %to convert eps images to pdf for use with pdf-latex
\usepackage{color}
\woctitle{5$^{th}$ Roma International Conference on Astroparticle Physics (RICAP-14)}
\begin{document}
\title{Direct Dark Matter Search with XENON100}
%%%      Or: with the XENON100 experiment
%%%      Recent results from XENON100 on WIMP and Axion Searches

\author{S.E.A. Orrigo\inst{1}\fnsep\thanks{Corresponding author \email{Sonja.Orrigo@ific.uv.es}}\fnsep\footnote{Present address: IFIC, CSIC-Universidad de Valencia, E-46071 Valencia, Spain} \\ \\ on behalf of the XENON Collaboration}

\institute{Department of Physics, University of Coimbra, Coimbra, Portugal}

\abstract{The XENON100 experiment is the second phase of the XENON program for the direct detection of the dark matter in the universe. The XENON100 detector is a two-phase Time Projection Chamber filled with 161 kg of ultra pure liquid xenon. The results from 224.6 live days of dark matter search with XENON100 are presented. No evidence for dark matter in the form of WIMPs is found, excluding spin-independent WIMP-nucleon scattering cross sections above $2 \times 10^{-45}$ cm$^2$ for a 55 GeV/c$^2$ WIMP at 90\% confidence level (C.L.). The most stringent limit is established on the spin-dependent WIMP-neutron interaction for WIMP masses above 6 GeV/c$^2$, with a minimum cross section of $3.5 \times 10^{-40}$ cm$^2$ (90\% C.L.) for a 45 GeV/c$^2$ WIMP. The same dataset is used to search for axions and axion-like-particles. The best limits to date are set on the axion-electron coupling constant for solar axions, $g_{Ae} < 7.7 \times 10^{-12}$ (90\% C.L.), and for axion-like-particles, $g_{Ae} < 1 \times 10^{-12}$ (90\% C.L.) for masses between 5 and 10 keV/c$^2$.}
\maketitle
\section{Introduction}
\label{intro}

Non-Barionic dark matter constitutes the 26.8 \% of the total energy and the 84.5\% of the total matter of the known universe. The nature of the dark matter is among the fundamental open questions in modern physics. One of the most favorite dark matter candidates are the Weakly Interacting Massive Particle (WIMPs), cold and not-charged exotic relics of the Big Bang that interact with the ordinary matter only through gravity and the weak interaction. Axions and Axion-Like Particles (ALPs) are other well-motivated candidates for cold dark matter, introduced by many extensions of the Standard Model of particle physics.

The XENON program aims at the direct detection of the dark matter in the universe using dual-phase Time Projection Chambers (TPCs) of increasing sensitivity \cite{Angle:2011th,Aprile:2011dd,Aprile:2012zx} filled with ultra pure liquid xenon (LXe). All the XENON experiments are located underground at the Laboratori Nazionali del Gran Sasso (LNGS) in Italy. The XENON100 experiment \cite{Aprile:2011dd}, having a total mass of 161 kg of LXe, is the second phase of the program and has set the most stringent limits on the WIMP-nucleon spin-independent \cite{Aprile:2012nq} and spin-dependent \cite{Aprile:2013doa} cross sections and on the axion-electron coupling \cite{Aprile:2014eoa}. The successor, XENON1T \cite{Aprile:2012zx,Rizzo:2014}, is a ton-scale TPC that will be commissioned in 2015 and it is expected to improve the sensitivity by two orders of magnitude.

\section{The XENON100 experiment}
\label{xenon100}

The XENON100 experiment \cite{Aprile:2011dd}, aimed primarily at the detection of WIMPs, is operated at LNGS since 2009 with outstanding stability and performances. The XENON100 detector is a position-sensitive two-phase (liquid-gas) TPC with a total mass of 161 kg of LXe, divided into a sensitive volume of 62 kg and 99 kg of active veto. Ultra pure LXe acts both as a target and detector medium, with the advantage of combining a high WIMP sensitivity with excellent self-shielding capability crucial for background suppression. The TPC is mounted in a double-walled vacuum-insulated $^{316}$Ti stainless steel cryostat, embedded in a passive radiation shield. Thanks to the careful material screening and selection \cite{Aprile:2011ru} and the detector design \cite{Aprile:2011dd}, XENON100 has achieved its design goal of reducing the electromagnetic background of two orders of magnitude in comparison to XENON10 \cite{Angle:2011th} while increasing the target mass by a factor of ten. The experimentally verified electromagnetic background is $< 5 \times 10^{-3}$ events/(keV$_{ee}$\footnote{keV electron-equivalent} kg day) in the low-energy region of interest for the dark matter search before signal discrimination \cite{Aprile:2011vb}.

\begin{figure}[!h]
\centering
\includegraphics[width=0.7\textwidth,clip]{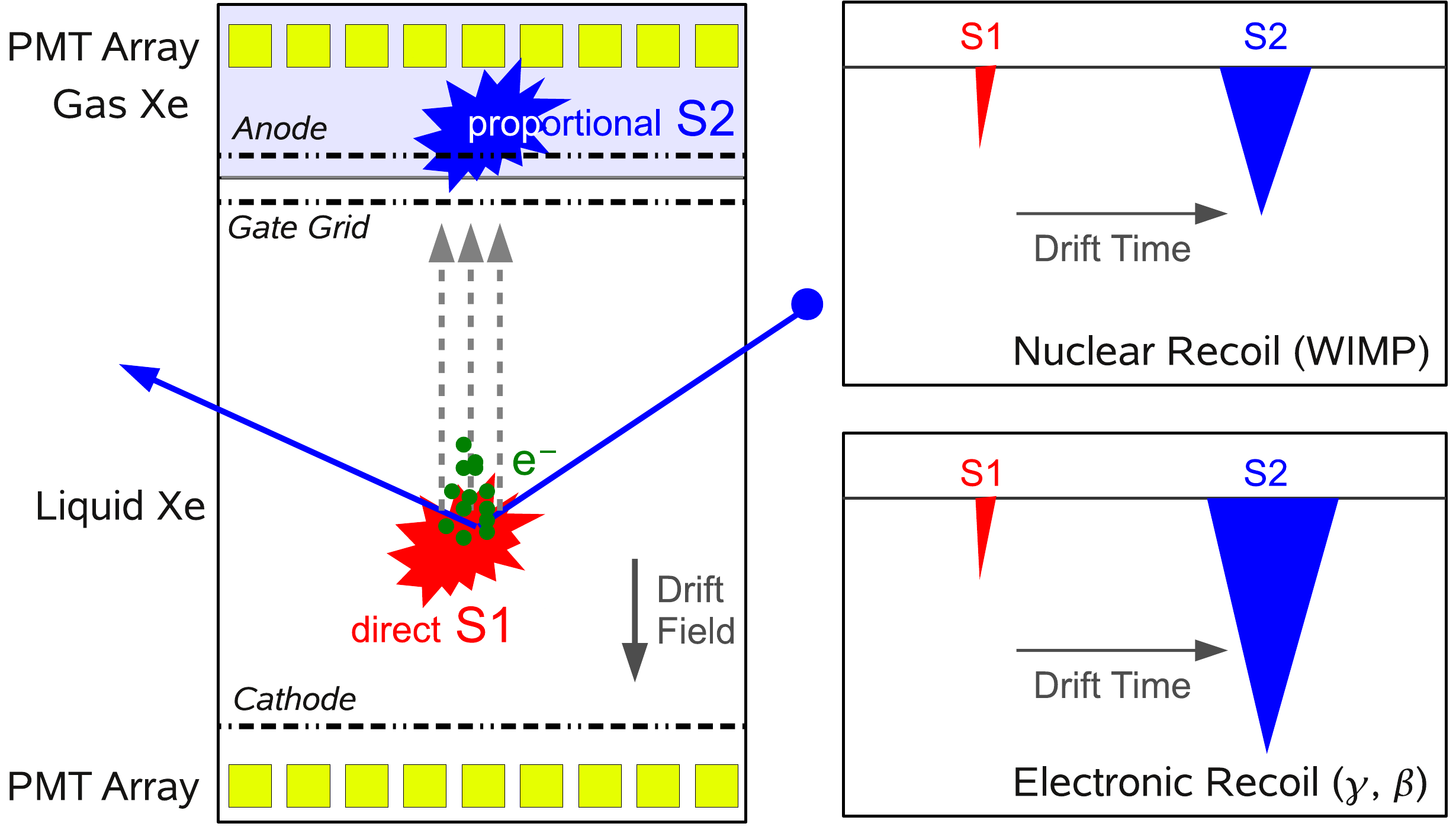}
\caption{Left: Working principle of the XENON TPC. Right: The signal ratio, $S2/S1$, allows discrimination between nuclear and electronic recoils.}
\label{fig-1}
\end{figure}

A scheme of XENON100 is shown in Fig. \ref{fig-1} (left). The TPC is equipped with low-radioactivity Hamamatsu R8520-06-Al 1” square photomultiplier tubes (PMTs), 98 of which are located in the gas phase ($top$ array). A second array of 80 PMTs is immersed in LXe below the cathode ($bottom$ array). Finally, 64 PMTs watch the active veto volume. A particle interaction in LXe produces direct scintillation light ($S$1 signal) detected by the bottom array. Ionization electrons are also produced; they are drifted towards the liquid-gas interface by the applied electric field and extracted into the Xe gas. A second signal, $S$2, is generated via proportional scintillation in the gas. The x- and y-coordinates of the interaction vertex are inferred from the proportional scintillation hit pattern on the top array. The time difference between the $S$1 and $S$2 signals, caused by the finite electron drift velocity in LXe, is proportional to the z-coordinate. Therefore the XENON100 TPC provides full 3-dimensional vertex reconstruction on an event-by-event basis. This property allows the definition of an ultra-low-background $fiducial~volume$ inside the target. Moreover, the $S2$/$S1$ ratio provides additional background discrimination because, as shown in Fig. \ref{fig-1} (right), it differs for electronic recoil signals (ER, from the $\gamma$ and $\beta$ backgrounds) and nuclear recoils (NR, from WIMPs and neutrons).

\section{Spin-independent and spin-dependent WIMP-nucleon couplings}
\label{wimps}
 
The interaction of the WIMPs with the target nuclei can be spin-independent (scalar coupling), which is possible with all the nuclei, and spin-dependent (axial-vector coupling). The latter is only possible if the WIMP carries spin and with odd mass isotopes, which have a non-zero nuclear spin ($^{129}$Xe and $^{131}$Xe in the case of XENON100). Recent results from the XENON100 experiment have put new constraints on the elastic, spin-independent \cite{Aprile:2012nq} and spin-dependent \cite{Aprile:2013doa} WIMP-nucleon cross sections.

224.6 live days of data were acquired between February 2011 and March 2012 \cite{Aprile:2012nq}. The data analysis is described in details in \cite{Aprile:2012vw}. The data were blinded below the 90\% ER quantile in $log_{10}$($S$2/$S$1)-space for $S1 <$ 160 PE. All the data quality cuts and topology requirements were designed and fixed before unblinding the signal region, according to the expected characteristics of the WIMP NR signal. The latter was calibrated using a $^{241}$AmBe source both before and after the science run. Calibrations of the low-energy ERs were taken regularly during the science run using $^{60}$Co and $^{232}$Th sources. A fit to the light yield $L_{y}$ measured with all the available calibration lines \cite{Aprile:2010um} determined the light yield $L_{y}$(122 keV$_{ee}$) = (2.28 $\pm$ 0.04) PE/keV$_{ee}$ at the applied drift field (530 V/cm). The detector response to single-electron charge signals has been characterized and used to infer crucial parameters for the experiment \cite{Aprile:2013blg}, showing an excellent sensitivity of XENON100 to small charge signals.

The NR energy scale has been inferred from the $S$1 signal as $E_{nr} = (S1/L_{y})(1/L_{eff})(S_{ee}/S_{nr})$ \cite{Aprile:2011hi}, where the relative scintillation efficiency $L_{eff}$ was extracted using all the existing direct measurements including \cite{Plante:2011hw}. The NR energy scale has been verified by a detailed Monte Carlo simulation of the whole detector, including the shield \cite{Aprile:2013teh}. The simulated $S$1 and $S$2 responses of XENON100 to NRs from an AmBe source have been compared to the experimental $S$1 and $S$2 spectra obtained by the AmBe calibration. Remarkable agreement has been found over the whole spectra (down to 2 PE in $S$1), reflecting the excellent understanding of the detector response to NRs at \% level.

Possible sources of background in the WIMP search region are NRs induced by neutrons and leaking ER events caused by the intrinsic radioactivity of LXe and the detector materials. The NR background was estimated by Monte Carlo simulations \cite{Aprile:2013tov} and the ER background from the calibration data \cite{Aprile:2011vb}. In the selected fiducial volume of 34 kg and for 99.75\% ER rejection, the background expectation in the WIMP search region is (1.0 $\pm$ 0.2) events \cite{Aprile:2012nq}, prediction verified in the high energy sideband (30–130 PE). After unblinding the data, 2 events survived to all the requirements for WIMP-induced single-scatter NRs, which do not constitute evidence for dark matter when compared to the expected background. Moreover, the profile likelihood analysis \cite{Aprile:2011hx} did not find any signal excess.

\begin{figure}[!ht]
	\begin{minipage}{0.48\linewidth}
		\centering
		\includegraphics[width=1\linewidth,clip]{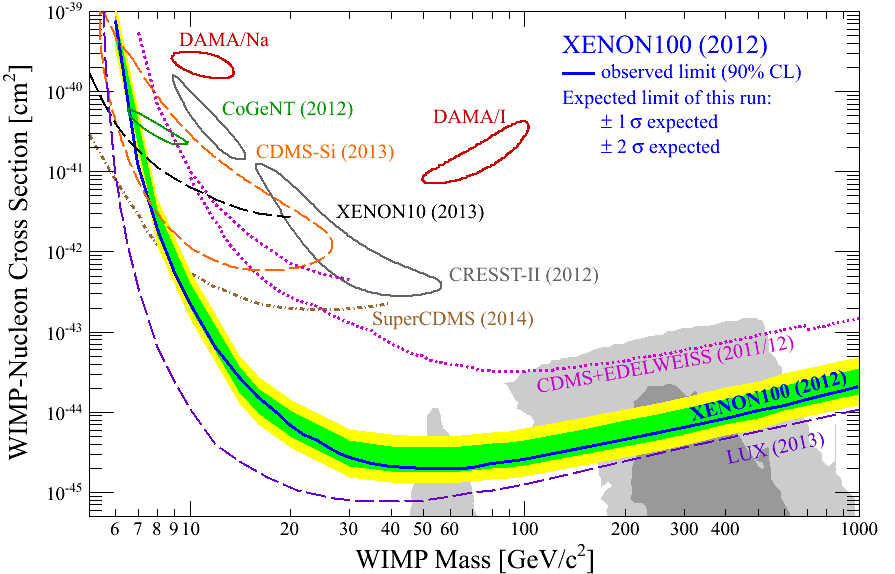}
    \caption{Spin-independent WIMP-nucleon elastic scattering cross section as a function of the WIMP mass. The XENON100 exclusion limit (90\% C.L.) is shown in blue; the green/yellow bands show the expected sensitivity (1$\sigma$/2$\sigma$) of this science run. Other experimental limits (90\% C.L.) and detection claims (2$\sigma$) are also shown for comparison.}
    \label{fig-2}
	\end{minipage}
	\begin{minipage}{0.48\linewidth}
		\centering
    \includegraphics[width=1\linewidth,clip]{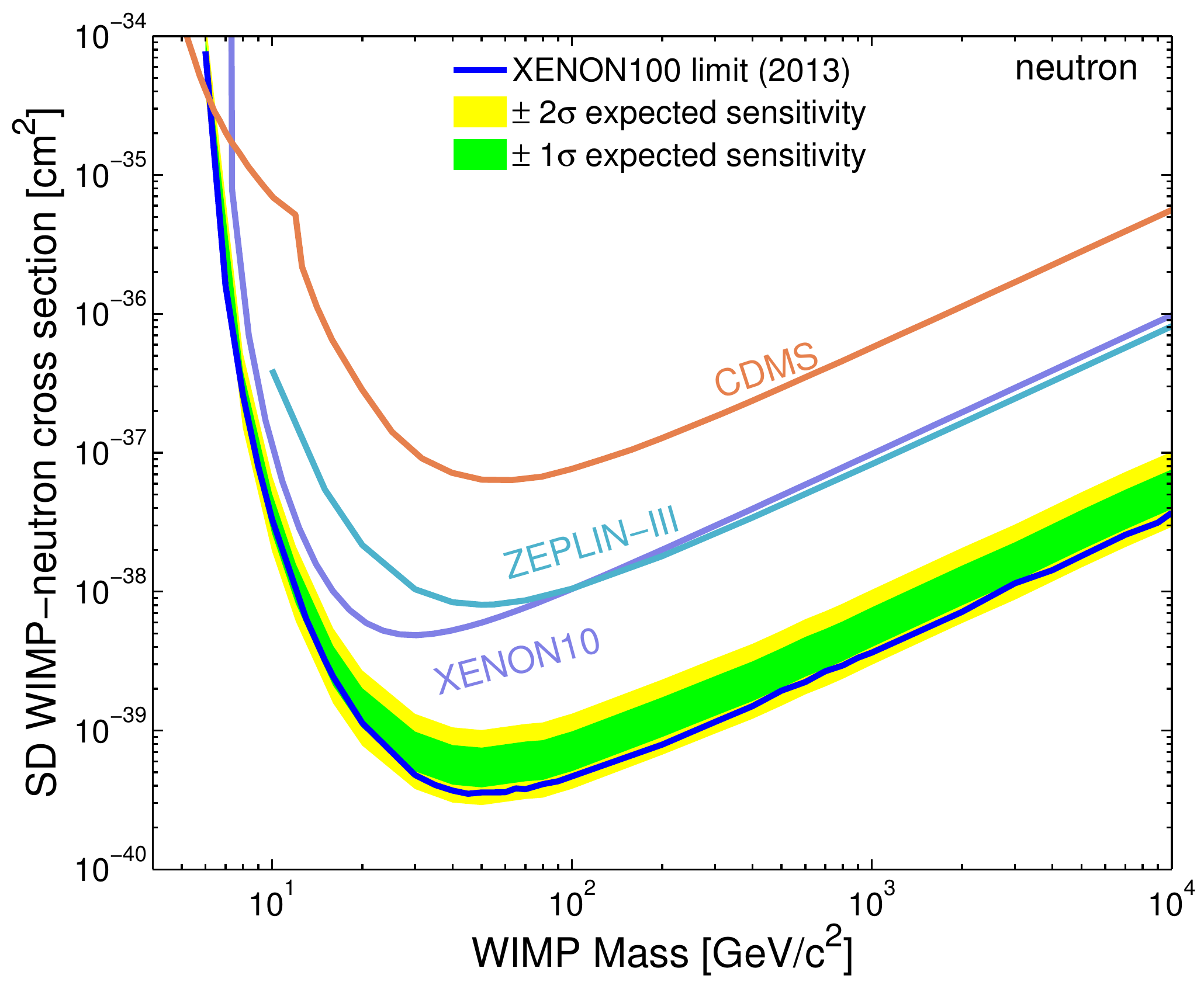}
    \caption{Spin-dependent WIMP-neutron elastic scattering cross section as a function of the WIMP mass. The XENON100 upper limit (90\% C.L.) is shown by the blue line; the 1$\sigma$/2$\sigma$ uncertainty on the expected sensitivity of the run is shown as a green/yellow band.}
    \label{fig-3}
	\end{minipage}
\end{figure}

A limit on the spin-independent WIMP-nucleon elastic scattering cross section has been calculated \cite{Aprile:2012nq} using the profile likelihood analysis \cite{Aprile:2011hx}, where the uncertainties in the background expectation and in the energy scale were profiled out. Poisson fluctuations in the generation of the PEs in the PMTs, which influence the energy resolution of the $S$1 signal, were also taken into account. It was assumed that the WIMPs are distributed in an isothermal halo with local density of $\rho_{\chi}$ = 0.3 GeV/cm$^{3}$, local circular velocity of $v_0$ = 220 km/s, and a Galactic escape velocity $v_{esc}$ = 544 km/s. The 90\% confidence level (C.L.) limit of Fig. \ref{fig-2} has a minimum at $2 \times 10^{-45}$ cm$^2$ for a 55 GeV/c$^2$ WIMP.

The same data and event selection as the spin-independent WIMP search, 224.6 live days for 34 kg of exposure, have been used to set a limit on the spin-dependent coupling, assuming that WIMPs couple predominantly to protons or neutrons \cite{Aprile:2013doa}. The nuclear model of \cite{Menendez:2012tm} was employed, which yields a far superior agreement between the calculated and measured spectra of $^{129}$Xe and $^{131}$Xe both in energy and in the ordering of the nuclear levels. The most sensitive limit on the pure WIMP-neutron cross section has been obtained above 6 GeV/c$^2$ (Fig. \ref{fig-3}), with a minimum at $3.5 \times 10^{-40}$ cm$^2$ (90\% C.L.) for a WIMP mass of 45 GeV/c$^2$. Even if the sensitivity on the pure WIMP-proton coupling is weaker because $^{129}$Xe and $^{131}$Xe have an unpaired neutron but even number of protons, the limit we get on the coupling to protons is also competitive \cite{Aprile:2013doa}.

\section{Solar axions and axion-like particles coupling to electrons}
\label{alps}

QCD axions and ALPs are light cold DM candidates which interact predominantly with atomic electrons in the medium and can couple to photons, electrons and nuclei. The coupling to electrons, $g_{Ae}$, can be tested through the $axio-electric~effect$. This process is analogous to the photoelectric effect, but with an axion/ALP playing the role of the photon: the axion ionizes an atom and it is absorbed, extracting an electron. Therefore axions and ALPs can scatter off the electrons of the LXe target, and so they can be searched for in XENON100 by looking at the ER events.

The same science run of 224.6 live days and 34 kg of exposure (see Sect. \ref{wimps}) has been used to perform the first axion searches with XENON100 \cite{Aprile:2014eoa}. The very low background and threshold of XENON100 are crucial for these analyses, leading to the first axion results from a dual-phase TPC. 

\begin{figure}[!ht]
	\begin{minipage}{0.48\linewidth}
		\centering
		\includegraphics[width=1\linewidth,clip]{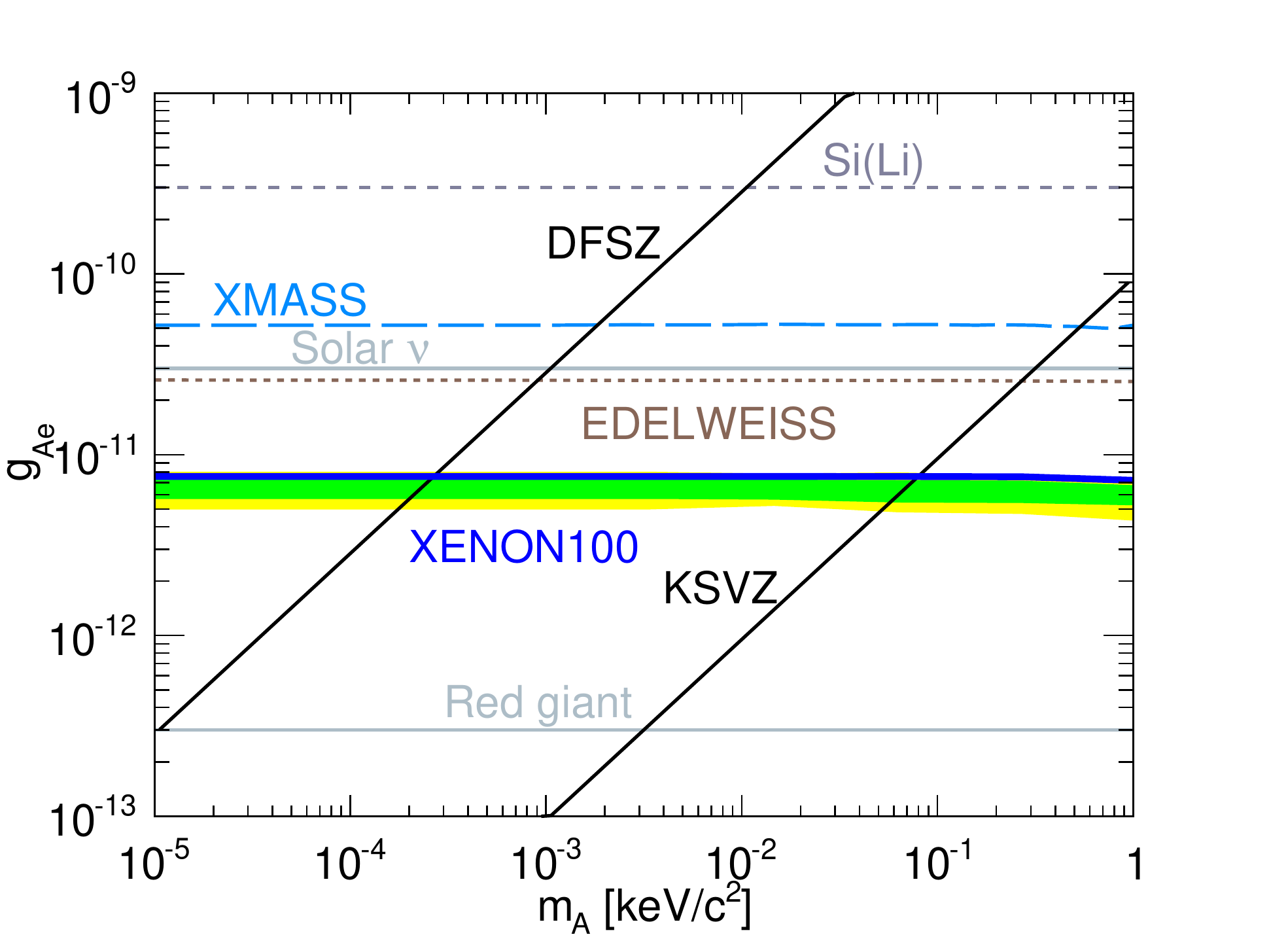}
    \caption{Axion-electron coupling constant for solar axions as a function of the axion mass. The XENON100 limit (90\% C.L.) is indicated by the blue line. The expected sensitivity, based on the background hypothesis, is shown by the green/yellow bands (1$\sigma$/2$\sigma$). Recent experimental constraints, astrophysical bounds and theoretical benchmark models are also shown for comparison.}
    \label{fig-4}
	\end{minipage}
	\begin{minipage}{0.48\linewidth}
		\centering
    \includegraphics[width=1\linewidth,clip]{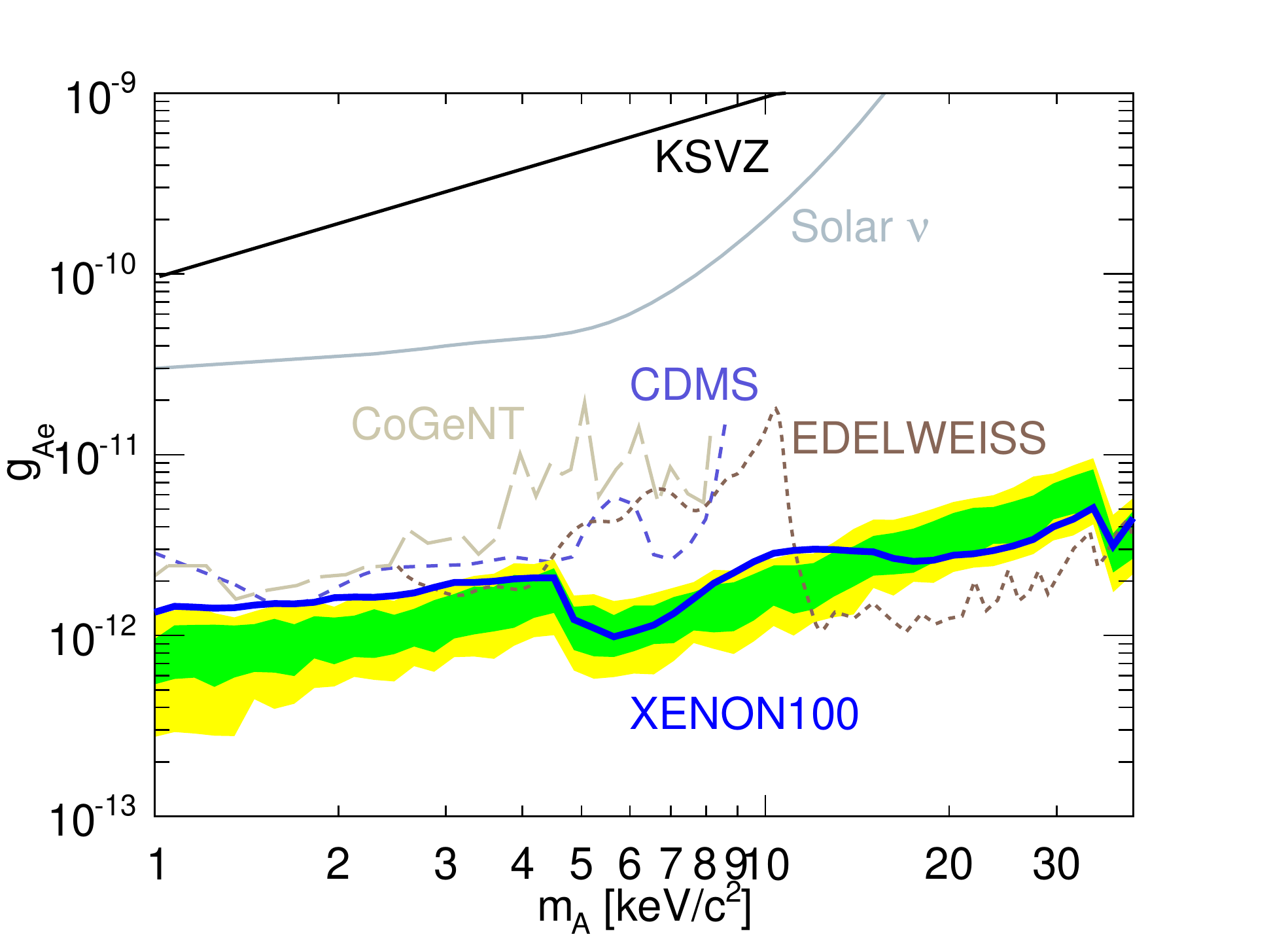}
    \caption{Axion-electron coupling constant for galactic axion-like particles as a function of their mass. The XENON100 limit (90\% C.L.) is shown by the blue line. The green/yellow bands indicate the expected sensitivity. Constraints set by other experiments, the solar neutrinos astrophysical bound and the benchmark KSVZ model are also shown.}
    \label{fig-5}
	\end{minipage}
\end{figure}

$Solar~axions$ are postulated to be produced in the Sun via Bremsstrahlung, Compton scattering, atomic recombination and atomic de-excitation. The event distribution for solar axions in XENON100 is expected to be a continuum \cite{Aprile:2014eoa}. A profile likelihood analysis gives no signal excess for the background-only hypothesis, thus the data are compatible with the background model. The best limit to date, shown in Fig. \ref{fig-4}, is set on the axion-electron coupling constant for solar axions with masses below 1 keV/c$^2$, $g_{Ae} < 7.7 \times 10^{-12}$ (90\% C.L.). In the frame of the DFSZ and KSVZ models, we exclude QCD axions heavier than 0.3 eV/c$^2$ and 80 eV/c$^2$, respectively.

$Galactic~ALPs$ in the keV range may have been generated via a non-thermal production mechanism in the early universe. The axion-electron coupling for galactic ALPs can be tested in XENON100 \cite{Aprile:2014eoa}, under the assumption that they constitute the whole abundance of dark matter in our galaxy. A monoenergetic peak at the axion mass is expected in the spectrum of galactic ALPs. The width of this signal is given by the energy resolution of the detector at the relevant $S$1 signal size. As for the solar axion search, the profile likelihood analysis shows no evidence for an ALP signal and so the data are consistent with the background hypothesis. The most sensitive limit, $g_{Ae} < 1 \times 10^{-12}$ (90\% C.L.), is obtained for galactic ALPs with masses between 5 and 10 keV/c$^2$ (see Fig. \ref{fig-5}).

\begin{thebibliography}{18}

\bibitem{Angle:2011th}
J.~Angle et~al. (XENON10 Collaboration), Phys.Rev.Lett. \textbf{107}, 051301
  (2011), \texttt{1104.3088}

\bibitem{Aprile:2011dd}
E.~Aprile et~al. (XENON100 Collaboration), Astropart.Phys. \textbf{35}, 573
  (2012), \texttt{1107.2155}

\bibitem{Aprile:2012zx}
E.~Aprile (XENON1T collaboration), Springer Proc.Phys. \textbf{C12-02-22}, 93
  (2013), \texttt{1206.6288}

\bibitem{Aprile:2012nq}
E.~Aprile et~al. (XENON100 Collaboration), Phys.Rev.Lett. \textbf{109}, 181301
  (2012), \texttt{1207.5988}

\bibitem{Aprile:2013doa}
E.~Aprile et~al. (XENON100 Collaboration), Phys.Rev.Lett. \textbf{111}, 021301
  (2013), \texttt{1301.6620}

\bibitem{Aprile:2014eoa}
E.~Aprile et~al. (XENON100 Collaboration), Phys.Rev. \textbf{D90}, 062009
  (2014), \texttt{1404.1455}

\bibitem{Rizzo:2014}
A.~Rizzo (on behalf of the XENON Collaboration), Proceedings of RICAP-14, EPJ
  Web of Conferences  (2014)

\bibitem{Aprile:2011ru}
E.~Aprile et~al. (XENON100 Collaboration), Astropart.Phys. \textbf{35}, 43
  (2011), \texttt{1103.5831}

\bibitem{Aprile:2011vb}
E.~Aprile et~al. (XENON100 Collaboration), Phys.Rev. \textbf{D83}, 082001
  (2011), \texttt{1101.3866}

\bibitem{Aprile:2012vw}
E.~Aprile et~al. (XENON100 Collaboration), Astropart.Phys. \textbf{54}, 11
  (2014), \texttt{1207.3458}

\bibitem{Aprile:2010um}
E.~Aprile et~al. (XENON100 Collaboration), Phys.Rev.Lett. \textbf{105}, 131302
  (2010), \texttt{1005.0380}

\bibitem{Aprile:2013blg}
E.~Aprile et~al. (XENON100 Collaboration), J.Phys. \textbf{G41}, 035201 (2014),
  \texttt{1311.1088}

\bibitem{Aprile:2011hi}
E.~Aprile et~al. (XENON100 Collaboration), Phys.Rev.Lett. \textbf{107}, 131302
  (2011), \texttt{1104.2549}

\bibitem{Plante:2011hw}
G.~Plante, E.~Aprile, R.~Budnik, B.~Choi, K.~Giboni et~al., Phys.Rev.
  \textbf{C84}, 045805 (2011), \texttt{1104.2587}

\bibitem{Aprile:2013teh}
E.~Aprile et~al. (XENON100 Collaboration), Phys.Rev. \textbf{D88}, 012006
  (2013), \texttt{1304.1427}

\bibitem{Aprile:2013tov}
E.~Aprile et~al. (XENON100 Collaboration), J.Phys. \textbf{G40}, 115201 (2013),
  \texttt{1306.2303}

\bibitem{Aprile:2011hx}
E.~Aprile et~al. (XENON100 Collaboration), Phys.Rev. \textbf{D84}, 052003
  (2011), \texttt{1103.0303}

\bibitem{Menendez:2012tm}
J.~Menendez, D.~Gazit, A.~Schwenk, Phys.Rev. \textbf{D86}, 103511 (2012),
  \texttt{1208.1094}

\end{thebibliography}
\end{document}